\documentstyle[11pt,paspconf]{article}

\def\lya{Ly$\alpha$~}
\def\kms{km~s$^{-1}$}
\newcommand{\lsim}{\ \raise -2.truept\hbox{\rlap{\hbox{$\sim$}}\raise5.truept
        \hbox{$<$}\ }}                  
\newcommand{\gsim}{\ \raise -2.truept\hbox{\rlap{\hbox{$\sim$}}\raise5.truept
        \hbox{$>$}\ }}

\input epsf

\begin{document}

\title{Scaling properties of the Lyman--$\alpha$ forest}

\author{S.~Savaglio}
\affil{European Southern Observatory, Karl-Schwarzschildstr. 2,
        D--85748 Garching bei M\"unchen, Germany}
\author{V.~Carbone}
\affil{Dipartimento di Fisica, Universit\'a della Calabria, I--87036 Roges 
di Rende (CS), Italy}

\begin{abstract}

We present some statistical
features of the large number of \lya absorption lines detected in high
redshift quasar spectra, obtained by using the multifractal 
approach. In the analysed sample of 12 QSO
sight--lines, 11 show scaling behaviour 
with a crossover between two distinct regimes: a non-homogeneous
regime at small scales and a homogeneous regime at large scales. 
The correlation length
shows a redshift dependence, suggesting
that the \lya forest can be an intermediate phenomenon between
a strongly inhomogeneous galaxy distribution
in the local Universe and
 a homogeneous initial mass distribution.

\end{abstract}


\section{Introduction}

The numerous Ly--$\alpha$ absorption lines seen in quasar spectra
can be considered a  very deep window on
the nature of the young Universe. All recent results obtained with high
resolution spectroscopy indicate  that the associated gas clouds
represent
most likely a consistent fraction of the dark side
of the baryonic matter physically connected with primeval galaxies.  
Clustering properties have been studied
basically using the two point 
correlation function. A positive signal was detected in high
resolution spectra  only for small scales (up to a few
hundred \kms, Webb 1987, Rauch et al. 1992, Chernomordik 1995,
Cristiani et al. 1996). 

The \lya absorption lines 
can be fitted by Voigt profiles in order to obtain  HI column
densities, Doppler widths and redshifts. At high redshift ($z \gsim 2$)
the  redshift evolution and HI column density distribution is well
reproduced by a double power law:

\begin{equation}
\frac{\partial^2 n}{\partial z \partial N_{\rm HI}} = A_o (1+z)^\gamma
N_{\rm HI}^{-\beta}
\end{equation}

\noindent
where $\gamma = 2 - 2.6$ and
$\beta = 1.4 - 1.7$. The important features of this kind of non
Gaussian distribution are the index of
the scaling laws and not the amplitudes at every scale and
 we analyse it using a mathematical tool
which is most suitable for their determination.

\begin{figure}[h]
\epsfxsize=11.6cm
\epsfysize=5.44cm
\centerline{\epsffile{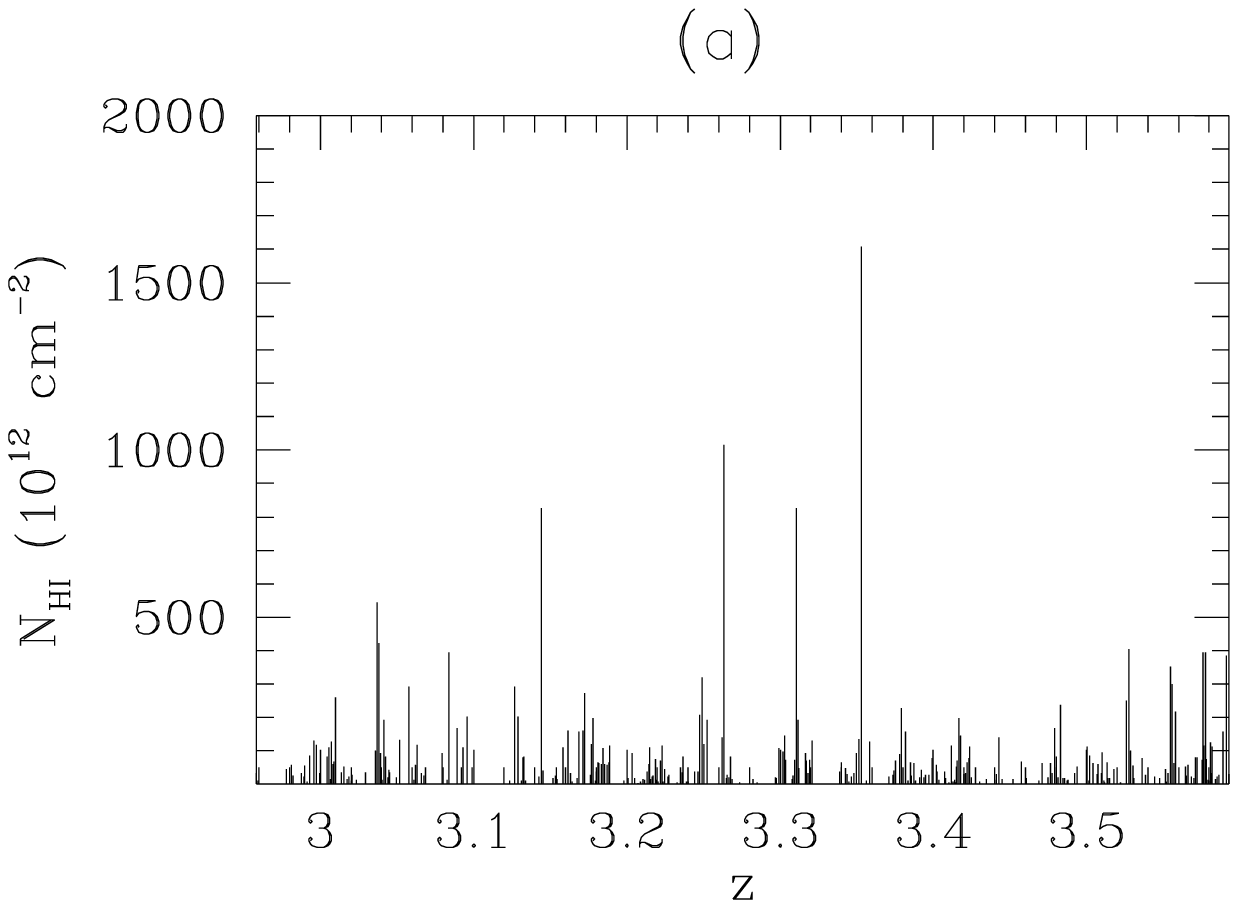}}
\epsfxsize=8cm
\epsfysize=5.6cm
\centerline{\epsffile{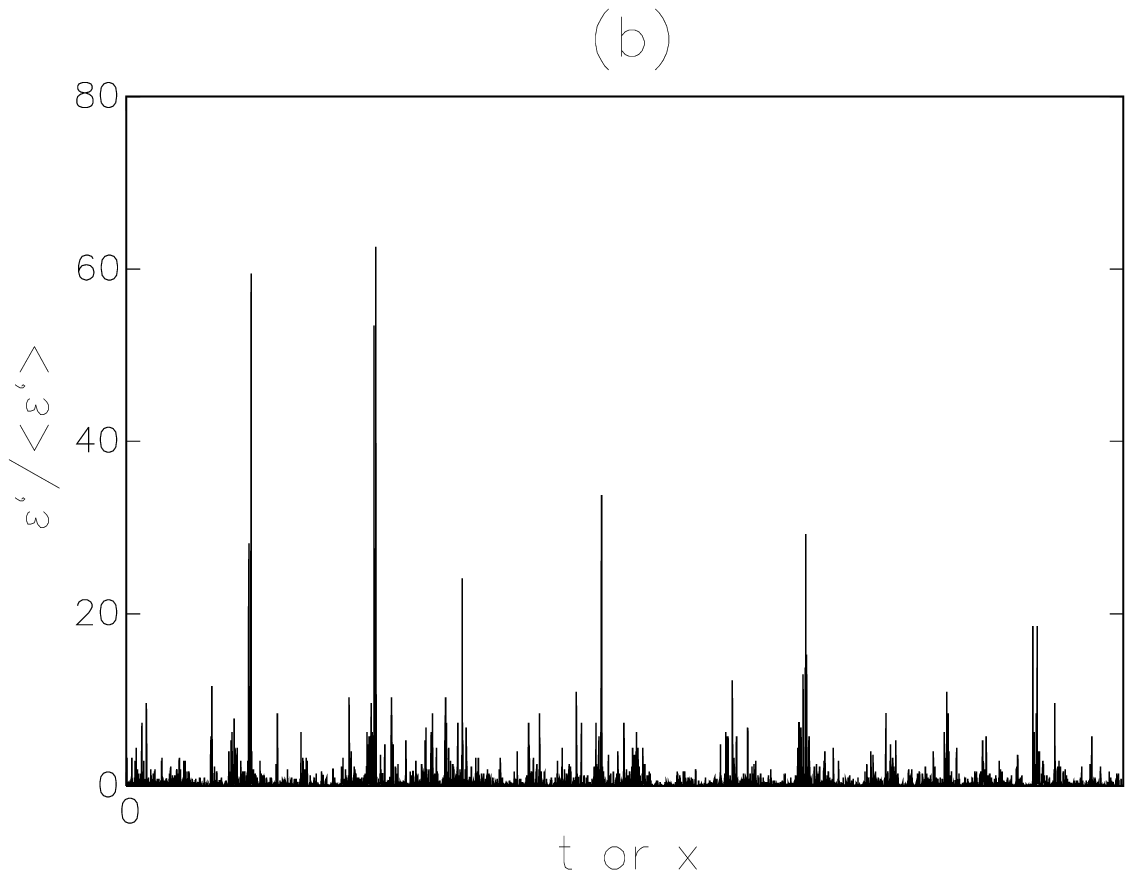}}
\caption[1]{\label{f1} Two examples of  multiplicative process.
(a) redshift distribution of HI column densities
in the spectrum of 
Q$0055-26$; (b) dissipation rate of the kinetic energy $\epsilon '$ 
in the atmospheric surface layer at a high Reynolds 
number (see Meneveau \& Sreenivasan, 1991).}
\end{figure}

\section{Turbulence and \lya clouds: The statistic of rare events}

In Fig.~\ref{f1} we show the visual similarity between the
distribution of two different physical quantities: the redshift
distribution of HI column
densities in a quasar spectrum and 
the kinetic energy dissipation in a fully developed turbulent  fluid.
The energy transfer, underlying the phenomenon shown in Fig.~\ref{f1}b, can be
described by means of a self--similar cascade with an associated multiplicative
process: a break--down of large--scale structures into small--scale ones,
each receiving a fraction of energy.
An analogous mechanism (for instance in a CDM scenario with an associated
``inverse cascade" of gravitationally confined structures) 
can leads to the phenomenon shown in
Fig.~\ref{f1}a. The result is an intermittent process, where rare
events (the localized peaks, or singularities, in the distribution)
have a higher probability to occur with
respect to a Gaussian process.

Every scaling process, like turbulence and HI column density 
distribution in the early Universe, can be treated in the
context of the fractal formalism or, more generally, the multifractal
formalism.
A multifractal is a scale-invariant distribution described
by a local exponent $\alpha$

\begin{equation}
P_i(r) \sim r^{\alpha}
\end{equation}

\noindent
where $P_i$ is the probability measure in the $i$--th interval of size
$r$ around the 
point $x$. In a multifractal, $\alpha$ depends on the position $x$.
To investigate the multifractal structure of the measure, we define
the generalized partition function of the
box--counting method (Paladin \& Vulpiani, 1987):

\begin{equation}
\chi^{(q)}(r) = \sum_i \left[P_i(r)\right]^q~,
\end{equation}

\noindent
where the sum is extended to all the subsets $i$ at a given scale $r$. The 
information relative to the multifractal structure can be recognized by 
calculating the generalized R\'enyi dimensions $D_q$ from the scaling law:

\begin{equation}
\label{equ2}\chi^{(q)}(r) \sim \left[r\right]^{(q-1)D_q}
\end{equation}

\noindent
For the way the partition function is defined, big values of $q$ emphasize the
scaling properties of overdense regions, while small values those 
of underdense regions. A multifractal structure 
is a non homogeneous fractal (with different $D_q$ at every $q$) where
the presence of clusters is enhanced by positive 
values of $q$ and that of voids  by negative values. 
 The two point correlation function
describes clustering at the first order only, while the 
infinite set of singularities, each being characterized by a different
fractal dimension $D_q$, 
describes clustering at every scale and has the important property to show
(in principle) the {\it entire hierarchy of clusters} (if any).  

\begin{figure}[h]
\epsfxsize=8.5cm
\epsfysize=7cm
\centerline{\epsffile{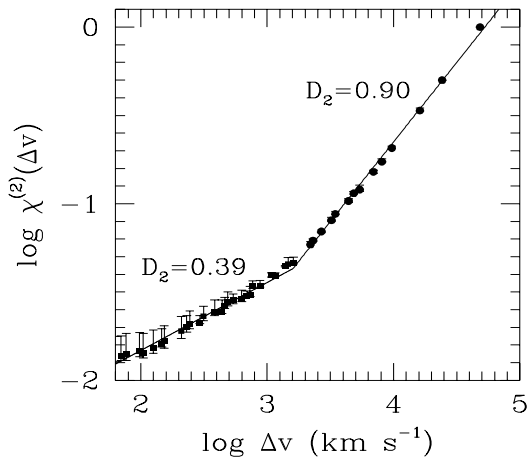}}
\caption[1]{\label{f2} The generalized partition function for $q=2$
as function of the redshift interval $\Delta v$ for the HI
column densities of Q$0055-26$.}
\end{figure}

\section{Multifractality in Q0055-26}

We have used in a
previous paper (Carbone \& Savaglio, 1996, hereafter CS96)
 the high resolution spectrum of Q$0055-26$ ($z_{em} = 3.65$,
Cristiani et al. 1995) in order to test the multifractality of the
\lya forest.
For each scale $\Delta v$ (between two redshifts $z_1$ and
$z_2$) we define a probability measure by 
dividing the redshift range into disjoint subsets $i$.
This measure $P_i(\Delta v)$ is defined as the total HI column 
density in the $i$--th subset characterized by a velocity separation 
$\Delta v$, normalized to the total column density in the spectrum. 
This can be related to the 
probability of occurrence of a certain amount of gas in the $i$-th box at a 
certain scale $\Delta v$. 

The results for $q=2$ 
are shown in Fig.~\ref{f2}. It is evident the presence of two regimes with
a linear relation and two different $D_2$ values. 
The separation between the two regimes occurs at $\log \Delta v \simeq
3.2$, 
which at a mean redshift of $z=3.305$
corresponds to a comoving scale of about
8 $h^{-1}$ Mpc.
Similar features are visible for higher values of $q$ up to $q = 4$
(CS96).

For comparison with the distribution of galaxies in the local
Universe, 
one can see Martinez et al. 1990,
Coleman \& Pietronero 1992, 
Borgani et al. 1994, Martinez \& Coles 1994 and Garrido et al. 1996.
In the multifractal analysis 
of the QDOT redshift survey of 2086
IRAS galaxies (dominated by spiral galaxies), Martinez \&
Coles (1994)
found multifractality and observed two regimes for different values of $q$.
For $q=2$, at small scales ($r<10~h^{-1}$ Mpc)
they found scaling properties with correlation dimension $D^{(3d)}_2 =
2.25$, which  in one dimension corresponds to a
fractal dimension of 0.25 ($D^{(3d)}_2 = D^{(1d)}_2+2$).
For large scales, the IRAS galaxies  reach
homogeneity. We notice that we are comparing the distribution of galaxies
of the local Universe with that of \lya clouds at high redshift. If the
change of regime occurs at similar comoving scales,  we conclude that 
\lya clouds have undergone a faster clustering evolution compared
to IRAS galaxies.
However multifractality in
galaxies is matter of controversy and it is strongly dependent on the
galaxy morphology.

The statistics of the \lya forest is poor
in comparison with galaxy surveys. One of the main problems
is thus to 
test the significance of the results. A first check 
has been presented in CS96. The observed distribution has been
compared with a set of 2000 ``fake'' distributions. In a following
work (Savaglio \& Carbone, in preparation),
new tests will be presented and sets of different simulations
will be compared to the observed distributions. As very preliminary
and crude results, we have seen that in 100 
simulations of the Q$0055-26$ sight--line,
53\% of the cases shows no scaling law, 37\% one single scaling law
with $0.58 < D2 < 0.73$.
In 10\% a very weak double scaling law, with $D_2$ which goes from about
0.6 at small scales to about 0.8 at large scales and a
correlation length  much smaller than the observed one.
 Even if a genuine multifractality in the \lya~clouds distribution
is evident, a richer sample of lines and comparison with simulations
would help to clarify this situation. 

\section{Scaling laws in a sample of QSO spectra}

\begin{figure}
\epsfxsize=14cm
\epsfysize=18cm
\centerline{\epsffile{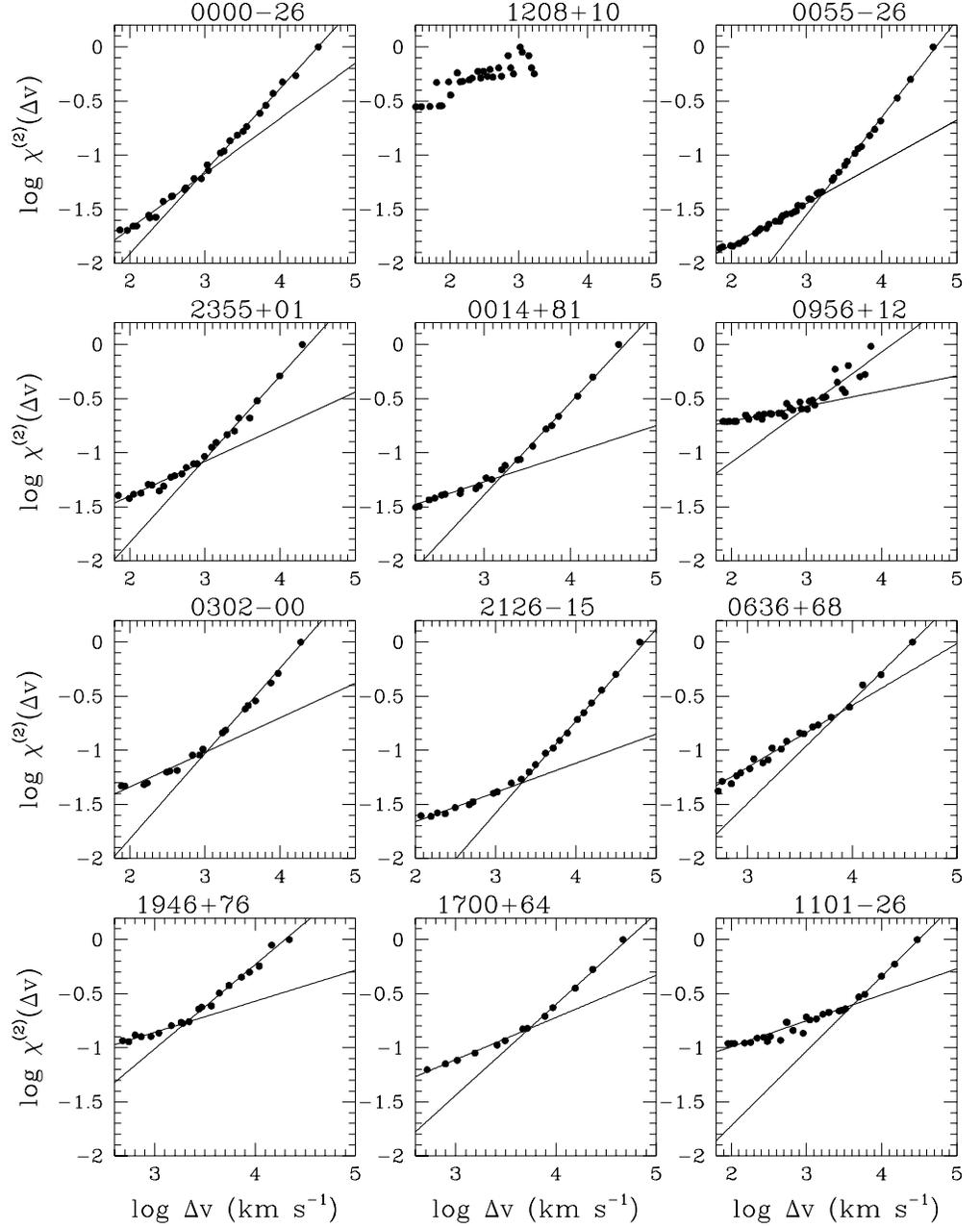}}
\caption[1]{\label{f3} The generalized partition function for $q=2$
as function of the redshift interval $\Delta v$ for the HI
column densities of 12 QSO sight--lines. The plots are in order of
decreasing redshift from the top--left to the bottom-right.}
\end{figure}

We applied the box--counting method to a
sample of QSO spectra. It represents part of the sample used
by Cristiani et al. (1996), with a total of 2412 lines for
 12 sight--lines. The
redshift coverage is  $1.85 \lsim z \lsim 4.12$ and
the resolution better than 15 \kms,  the best available
 for these sources.
The lower limit for the redshift is imposed by the lack of statistics of 
HST high resolution data. 
We restricted our analysis to high resolution spectroscopy in order
to minimize the problem of line blanketing (the confusion of lines)
which at high redshift can dramatically affect  the analysis.

\begin{figure}[h]
\epsfxsize=8.8cm
\epsfysize=7cm
\centerline{\epsffile{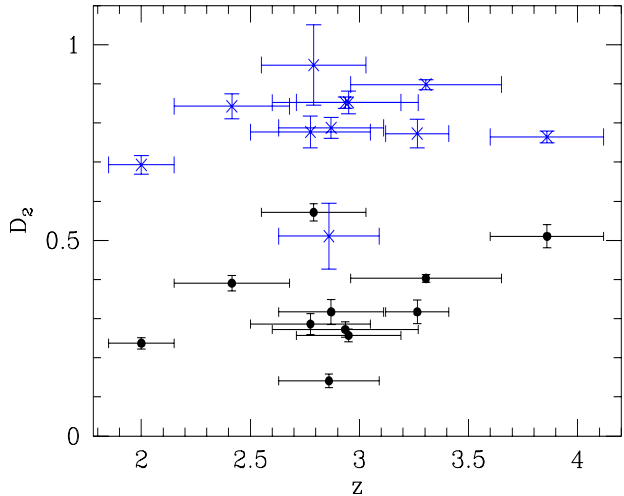}}
\caption[1]{\label{f4} Generalized dimension $D_2$ as function of
redshift for the 11
\lya forests in the two regimes of large scales (cross) and
small scales (filled circles).}
\end{figure}

\begin{figure}[h]
\epsfxsize=8.5cm
\epsfysize=6.5cm
\centerline{\epsffile{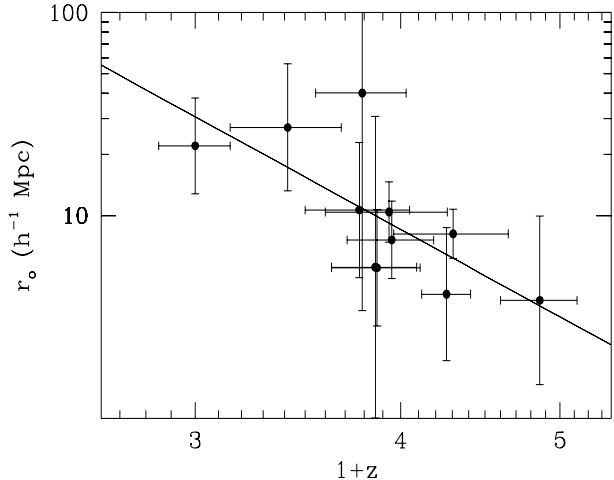}}
\caption[1]{\label{f5} Redshift distribution of the correlation
lengths (the scale at which the
partition function changes slope). 
The straight line is the linear regression of the points. }
\end{figure}

The partition functions with 
$q=2$ for the 12 \lya forests are shown in Fig.~\ref{f3}.
We confirm the presence of two regimes  in almost all of them,
except for one object ($1208+10$), where we do not see any clear 
scaling law in the distribution. For two objects ($0956+12$ and $0636+68$) 
the determination of the two scales is particularly difficult.
For the remaining 9 objects, two slopes are clearly visible.
In the plot of $D_2$ as function of the
mean redshift for the two different
scales (Fig.~\ref{f4})
there is no correlation for large scales, with a mean value
of  0.8. This value is an indication that homogeneity has been reached in the
sample. For small scales there is a clear deviation from homogeneity.
A weak  correlation with redshift of $D_2$, 
being smaller for lower redshifts, can also be noticed. 
A more clear redshift evolution  is shown by the distribution of the
correlation lengths (Fig.~\ref{f5}). A simple fit with a power law
gives:

\begin{equation}
r_o \propto (1+z)^{-4.5}
\end{equation}

\noindent
These results suggest
a picture where an initial homogeneous distribution of gas clouds or
mass in the Universe is broken by process of fragmentation  (in a Cold
Dark Matter scenario) or of aggregation
of matter around some singularities (in the Hot Dark Matter scenario).
The ultimate fate of both the processes is a highly intermittent 
distribution like
that shown in Fig.~\ref{f1}a. 

\section{Conclusions}

The work presented here is in progress. A more extended analysis,
with a full description of different methods testing the
stability of the results, will
be presented elsewhere.
Even if the main
problem is the lack of statistics, we can firmly conclude
that multi-scaling analysis of \lya forests 
is a very promising approach to the study of the large
scale structure of the Universe at high redshift and its evolution.
This has to be regarded as a parallel and complementary 
point of view with respect 
to the study of the  galaxy distribution.

The two point correlation function analysis
can be replaced by different statistics which are more suitable to
describe highly inhomogeneous distributions. 
\lya clouds have shown
scaling laws for much larger scales with respect to previous analysis,
around 10 $h^-1$ Mpc 
in comoving distance at redshift of about $3 - 3.5$.
For larger scales, we have no indication against a homogeneous
distribution.

The multifractal properties of a sample of 11 quasars show 
evolution with redshift. 
In particular both the amplitude and the strength of the
multifractality decrease with redshift, which is what one expects to
see  in a Universe where gravitational clustering gives rise to larger,
correlated structures.

These results open the possibility to new scenarios were a local galaxy
distribution strongly inhomogeneous up to very large scales, is
compatible with a homogeneous initial mass distribution.

\acknowledgments

We are grateful to L.~Amendola for helpful discussions. We also thank
C. Meneveau for kindly providing us with the postscript file of 
Fig.~\ref{f1}b, and S. Cristiani, S. D'Odorico, V. D'Odorico, A. Fontana
 and E. Giallongo for making the data of the ESO Key Programme available.

\end{document}